\documentclass[aip,apl,amsmath,reprint]{revtex4-1}
\usepackage{graphicx}
\usepackage{upgreek}

\begin{document}

\title{Femtosecond laser pulse train interaction with dielectric materials} 

\author{O. Dematteo Caulier}
 \email{dematteo@celia.u-bordeaux1.fr}
 \author{K. Mishchik} 
\author{B. Chimier}%

\author{S. Skupin}

\author{A. Bourgeade}

\affiliation{%
Univ.~Bordeaux - CNRS - CEA, Centre Lasers Intenses et Applications, UMR 5107, 33405 Talence, France}%
\author{C. Javaux L\'eger}

\author{R. Kling}
\affiliation{
ALPHANOV, rue Fran\c cois Mitterand, 33400 Talence, France}

\author{C. H\"{o}nninger}
\affiliation{
AMPLITUDE SYSTEMES, 11 avenue de Canteranne, Cit\'e de la Photonique, 33600 Pessac, France}
\author{J. Lopez}

\author{V. Tikhonchuk}

\author{G. Duchateau}
\affiliation{%
Univ.~Bordeaux - CNRS - CEA, Centre Lasers Intenses et Applications, UMR 5107, 33405 Talence, France}%

\date{\today}

\begin{abstract}
We investigate the interaction of trains of femtosecond microjoule laser pulses with dielectric materials by means of a multi-scale model. Our theoretical predictions are directly confronted with experimental observations in soda-lime glass. We show that due to the low heat conductivity, a significant fraction of the laser energy can be accumulated in the absorption region. Depending on the pulse repetition rate, the material can be heated to high temperatures even though the single pulse energy is too low to induce a significant material modification. Regions heated above the glass transition temperature in our simulations correspond very well to zones of permanent material modifications observed in the experiments. 
\end{abstract}


\maketitle 
The use of femtosecond (fs) laser pulses for modifications of transparent materials is nowadays a well established technique with many applications in science and engineering. Material modification caused by a pulsed fs laser, i.e.\ a train of fs pulses, can be controlled by numerous experimental parameters such as repetition rate~(RR)~\citep{miyamoto_fusion_2007}, incident energy~\citep{peng_control_2013}, or focusing conditions~\cite{Zhu20052153}. Thus, ultrashort intense laser pulses are a versatile, highly adaptable processing tool for micro-machining of various materials~\cite{Cheng201388}, including human tissues in the context of ophthalmic surgery~\cite{Kymionis}. 

When a fs laser pulse is focused into a dielectric material to intensities exceeding a few TW/cm$^2$, a considerable number of electrons undergo photo- and impact-ionization processes~\cite{keldysh_ionization_1965}: electrons from the valence band (VB) are promoted to the conduction band (CB)~\cite{brouwer_excitation_2014}. Once the carrier density in the CB becomes non-negligible, these electrons strongly influence the pulse propagation dynamics, and the laser energy is efficiently absorbed in the near focal-region~\cite{gamaly_transient_2014}. At the end of the irradiation process, typically after a hundred fs, the electrons in the CB transfer their energy to the lattice through collisional processes, leading to an increase of the material temperature in the focal volume~\citep{burakov_spatial_2007}. On longer time scales, this energy is transferred toward the surrounding cold matter through heat diffusion \cite{bulgakova_pulsed_2014}. 
Provided that the amount of laser energy absorbed in the focal volume is sufficiently large, permanent material modifications are induced. This can be achieved either with a single fs microjoule laser pulse, or by using cumulative heating techniques~\cite{Schaffer03}, i.e., multiple pulses or pulse trains. Due to the low heat diffusivity in dielectric materials ($\sim10^{-3}$~cm$^2$/s), the cooling time over the typical size of the focal spot (a few microns) is of the order of $\sim10$~$\upmu$s. Therefore, by adjusting the time interval between consecutive laser pulses appropriately (kHz--MHz RR), one may accumulate the laser energy in the absorption region and reach very high temperatures, even if the single pulse energy is too low to induce a permanent material modification~\cite{couairon_femtosecond_2007}.

Several theoretical investigations concerning the interaction of fs laser pulses with dielectrics can be found in the literature. For instance, Brouwer~\textit{et al.}~\cite{brouwer_excitation_2014} provided a detailed study of fs pulse propagation and electronic excitation during the interaction. Bulgakova~\textit{et al.}~\cite{bulgakova_laser-induced_2010} evaluated the energy deposition by a single fs pulse, in order to model the heat source in subsequent simulations of thermal accumulation. Eaton~\textit{et al.}~\cite{eaton_heat_2005} or Jamshidi-Ghaleh~\textit{et al.}~\citep{jamshidi-ghaleh_laser_2006}, simulated heat accumulation from pulse trains by using phenomenological expressions for the deposited energy obtained from experimental observations. Here, for the first time, we confront rigorous quantitative theoretical predictions of permanent material modifications induced by a train of fs laser pulses with experimental results. 

In this paper, we develop a multi-scale model covering the key physical processes eventually leading to the observed permanent modification of the glass through heat accumulation. 
This modification may be caused by, e.g., structural change, creation of color centers, etc. In our model, we simply assume that permanent modification occurs for lattice heating above a given threshold temperature.
Our approach is based on separation of the different time-scales involved, i.e., fs or ps for pulse propagation and energy deposition, and $\upmu$s or ms for the thermal processes. First, we simulate the highly nonlinear fs pulse propagation accounting for multi-photon and collisional ionization as well as the optical Kerr effect. In a second step, we extract a map of the deposited energy density in the focal volume. This energy density map is then used as a source term in a heat equation in order to compute the ms evolution of the thermal distribution. 
The model is evaluated for trains of up to 500 pulses with microjoule energies and kilohertz RRs focused into soda-lime glass to investigate the heat accumulation regime. Results are directly confronted with experiments for quantitative comparison. Good agreement is obtained for threshold temperatures located in between strain and annealing point of the material. 

Let us start with a more detailed description of fs laser pulse propagation in transparent media, thus considering the fs-ps timescale. To this end, we resort to an extended nonlinear optical Schr\"odinger equation accounting for self-focusing due to the optical Kerr effect and light defocusing via laser-generated conduction electrons~\cite{berge_ultrashort_2007}. For the considered case of a linearly polarized light and moderate focusing conditions, the scalar and paraxial approximations are justified. Moreover, by comparing to a more complete unidirectional model~\cite{berge_ultrashort_2007} we checked that higher order effects like self-steepening can be neglected. 
Thus, the slowly varying complex optical field envelope $\varepsilon(r,\bar{t},z)$ is governed by the following equation
\begin{equation}
  \begin{split}
    \partial_z {\varepsilon} & = \dfrac{i}{2k_0}\dfrac{1}{r}\partial_r r \partial_r
{\varepsilon}- i\dfrac{k''}{2}\partial^2_{\bar{t}}{\varepsilon} 
  -\dfrac{\sigma}{2}N_e{\varepsilon} + i\dfrac{\omega_0 n_2}{c} I{\varepsilon} 
  \\  & \quad  -i\dfrac{k_0}{2n_0^2N_c}N_e{\varepsilon}    - \dfrac{E_gW_{PI}\left(N_{nt}-N_e \right)}{2I} {\varepsilon}   ,
 \label{krausz_radial}
  \end{split}
\end{equation}
where $n_0$ and $n_2$ are the linear and nonlinear refractive index, respectively, $k''$ corresponds to the second order group velocity dispersion in the material, $N_c$ is the critical electron density in the conduction band, $E_g$ is the band gap, $I(r,\bar{t},z)$ is the laser pulse intensity, where the optical field envelope ${\varepsilon}$ is normalized such that $I=|{\varepsilon}|^2$. The time $\bar{t}$ is the retarded time in a frame moving with the group velocity of the pulse at center frequency $\omega_0$ and wave number $k_0=n_0\omega_0/c$. The laser energy absorption by conduction band electrons is described by the Drude model with the cross section $\sigma=k_0 e^2 \tau_c/[n_0^2 \omega_0 \varepsilon_0 m_e (1+\omega_0^2\tau_c^2)]$, where $\tau_c$ is an effective electron collision time.
The evolution of the electron density in the conduction band $N_e(r,\bar{t},z)$ reads
\begin{equation}\label{n_evol}
  \partial_{\bar{t}} N_e  = W_{PI}\left(N_{nt}-N_e \right)+ N_e \dfrac{\sigma I}{E_g}- \dfrac{N_e}{\tau_{{rec}}},
\end{equation}
where $N_{nt}$ is the initial density of valence electrons, $W_{PI}$ is the photo--ionization rate derived from Keldysh formula~\cite{keldysh_ionization_1965}, and $\tau_{rec}$ is a typical recombination time.

Solving Eqs.~\eqref{krausz_radial} and \eqref{n_evol} gives the spatial distribution of laser energy density absorbed by the CB electrons during the whole propagation of a single laser pulse. This deposited energy density $U(r,z)$ is assumed to be fully transferred to the lattice much faster than the characteristic heat diffusion time. Thus, irradiation by each laser pulse increases the lattice temperature $T(r,t,z)$ by $T_L(r,z)=U/\rho C_v$, where $\rho$ and $C_v$ are the density and heat capacity of the material, respectively. The spatial and temporal evolution of $T$ is then determined by
\begin{equation}\label{heatdif}
   \dfrac{\partial T}{\partial t} = D {\nabla}^2 T + \sum\limits_{j=0}^{N-1} T_L \, \delta (t - j\Delta t).
\end{equation}
Here, the first term on the right-hand-side describes the temperature diffusion process with the thermal diffusion coefficient $D$, and the last term corresponds to the heat source caused by a train of $N$ pulses separated by $\Delta t$. Because the expected material modifications are rather weak, we neglect any temporal variation of the diffusion coefficient $D$ and the source term $T_L$. Both quantities are evaluated at room temperature.

By solving Eq.~\eqref{heatdif}, the lattice temperature evolution of the material irradiated by a fs pulse train is obtained. If the thermal diffusion process in Eq.~\eqref{heatdif} is slower than the heating process, the increase in the temperature can be significant for long pulse trains containing several hundreds of pulses. We assume a permanent material modification in regions where the temperature becomes larger than a given threshold temperature. This condition defines both the size and the shape of the material modification with respect to the laser pulse train parameters.

Simulations are performed for the soda-lime glass irradiated by trains of 500 pulses with 290~fs duration at 1030~nm wavelength and 1.3~$\upmu$J single pulse energy. Pulse parameters and focusing conditions are chosen corresponding to the experiments presented below, where the position of the geometrical focus is about 500~$\mu$m below the sample surface inside the bulk material. All relevant material parameters are summarized in Tab.~\ref{coeff_fs}.

\begin{table}
  \begin{tabular}{|c|c||c|c|}
    \hline
    $n_0$ & 1.5134 & $\sigma$ [cm$^2$]& $1.05 \times 10^{-17}$ \\
    \hline
    $n_2$ [cm$^2$/W]& $3.86 \times 10^{-16}$ &$\tau_c$ [fs]&  1.83 \\
    \hline
    $E_g$ [eV]& 3.9 &$\tau_{ rec}$ [fs] & 150 \\
    \hline
    $k''$ [fs$^2$/cm]& 286 & $\rho$ [g/cm$^{3}$] &   2.44 \\
    \hline
    $N_c$ [cm$^{-3}$]& $1.05 \times 10^{21}$  & $D$ [cm$^2$/s]& 0.72 \\
    \hline
    $N_{nt}$ [cm$^{-3}$]&$2.1 \times 10^{22}$ &$C_v$ [J/g.K ] & $5.97 \times 10^{-3}$ \\
    \hline
  \end{tabular}
\caption{Material parameters used to simulate femtosecond pulse train interaction with soda-lime glass\citep{jewell_thermooptic_1993, sun_diagnose_2006, sodalime}.}\label{coeff_fs}
\end{table} 

\begin{figure*}
  \includegraphics[width=1.4\columnwidth]{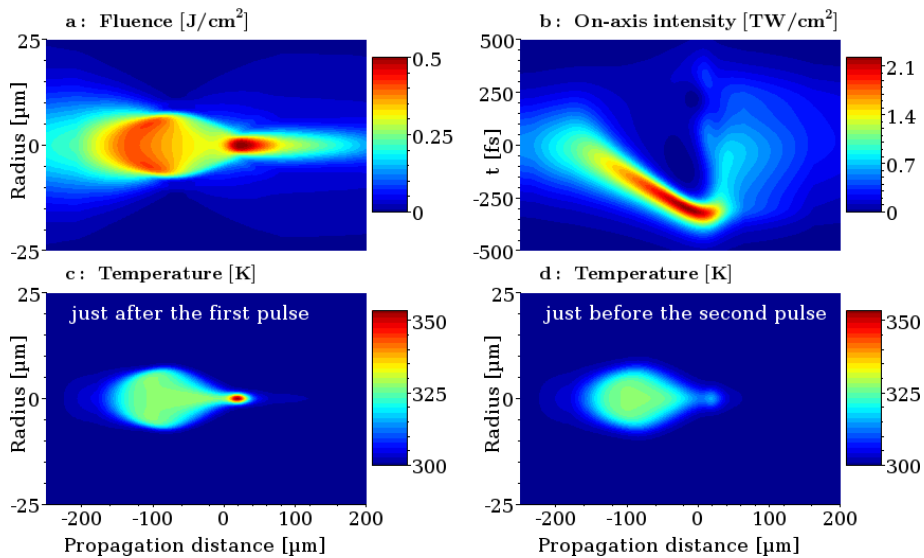} 
  \caption{(Color online) Fluence (a) and on-axis temporal profile (b) versus propagation distance for a single pulse  (1.3~$\mu$J, 290~fs at 1030~nm) propagating in soda-lime glass according to Eqs.~\eqref{krausz_radial} and \eqref{n_evol}, for an initial beam diameter of 5~mm and a focal length of 20~mm (in air).  The deposited energy density by such single pulse is directly proportional to the temperature distribution $T_L$ transferred to the lattice (c). The evolved temperature distribution according Eq.~\eqref{heatdif} after 2~$\upmu$s, just before the second pulse arrives, is shown in (d). The laser pulse propagates from left to right, with geometrical focus at zero.}
\label{doublefignrj+temp} 
\end{figure*}

Figure~\ref{doublefignrj+temp}(a,b) presents spatial and temporal dynamics of the single fs pulse propagation in soda-lime glass obtained from Eqs.~\eqref{krausz_radial} and \eqref{n_evol}. The evolution of the beam fluence shown in panel (a) confirms a focal spot size of a few microns. The on-axis temporal evolution shown in panel (b) is typical for focused fs low-energy pulses: a short (here $\sim50$~fs) intense spike at the pulse front is responsible for a significant free electron production, which in turn defocuses the trailing part of the pulse.

From the single fs pulse propagation simulation, we extract the deposited energy density $U$ in order to construct the temperature distribution $T_L$ transferred to the lattice, i.e., the source term in Eq.~\eqref{heatdif}. In the simulation, 60~\% of the laser energy is absorbed in the material, which is close to 55~\% measured experimentally. The main part (95~\%) of the absorbed laser energy is deposited in the material located before the focal point, even though the absorbed energy density is maximum at the focus (see Fig.~\ref{doublefignrj+temp}(c)). This particular spatial distribution is induced by the promotion of electrons to the CB during the propagation, because these electrons absorb the laser energy very efficiently. After the focal point, the intensity rapidly decreases and becomes too low to create a significant carrier density in the CB. 

Our simulations results of the single fs laser pulse propagation show that the maximum temperature reaches $T=360$~K at the focus (see Fig.~\ref{doublefignrj+temp}(c)), which is much smaller than the strain temperature $T_s=800$~K or annealing temperature $T_a=830$~K in the soda-lime glass~\cite{sodalime}.
Therefore, we expect no material modification from a single fs laser pulse. However, Fig.~\ref{doublefignrj+temp}(d) reveals that 2~$\upmu$s later, i.e., just before the second pulse arrives, the temperature in the interaction zone is still significantly elevated. In fact, the heat transfer in the surrounding material is slow, only the spatially very narrow temperature peak at the focus has decreased to some extend, and the temperature in the overall interaction zone is still well above the initial 300~K. It would take more than 100~$\upmu$s to relax to temperatures below $305$~K in the whole volume after a single pulse interaction. Thus, the interaction volume will get heated to higher and higher temperatures by each pulse of the incident fs pulse train, and a thermal accumulation process takes place.

\begin{figure}
  \includegraphics[width=0.9\columnwidth]{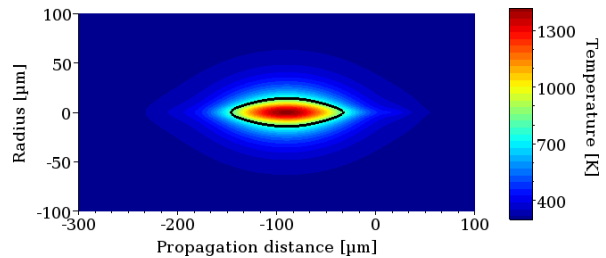} 
  \caption{(Color online) Spatial distribution of the temperature in soda-lime glass  after 1~ms of irradiation by 500 pulses at 500~kHz RR. The zone with temperature above $T_s$ is encircled by the black line. The laser pulses propagate from left to right.}
\label{fig2} 
\end{figure}

The lattice temperature obtained after 1~ms (accumulation over 500 pulses at a 500~kHz RR) is shown in Fig.~\ref{fig2}. The interaction zone is located about 100~$\mu$m before the focal point. The radial dimension of the heated volume in Fig.~\ref{fig2} is much larger than the one affected by the single fs pulse (c.f.~Fig.~\ref{doublefignrj+temp}(c) and (d)). The maximum temperature after 1~ms is close to 1350~K, i.e., well above both $T_s$ and $T_a$, and a permanent material modification can be expected in the region where $T>T_s$, encircled by the black line in Fig.~\ref{fig2}. 

\begin{figure}
 \includegraphics[width=0.6\columnwidth]{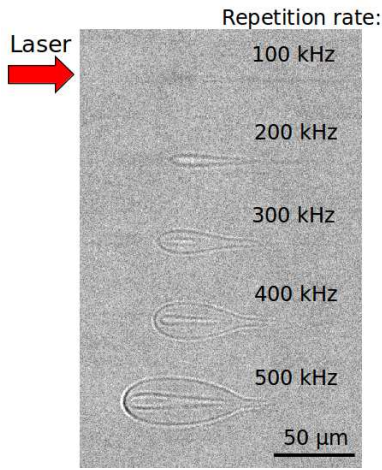}
 \caption{Transverse optical microscopy images of the permanent material modifications induced in soda-lime glass for 500 pulses for various RRs.}
 \label{ResExpRR}
 \end{figure}

To compare our numerical predictions with experimental observations, pulse trains produced by an Yb-doped fiber laser (Satsuma, Amplitude Syst\`emes) were focused into a sample of a bulk soda-lime glass with a x10 Mitutoyo objective lens ($\rm{NA}=0.26$; $f=20$~mm). 
Figure~\ref{ResExpRR} presents transverse images of the interaction volume after irradiation with 500 pulses for various RRs. We report no measurable modification of the matter for 100~kHz RR; permanent modifications of the material clearly appear for RRs larger than 200~kHz. The size of the structure increases with the RR towards the surface of the sample.

\begin{figure}
  \includegraphics[width=0.8\columnwidth]{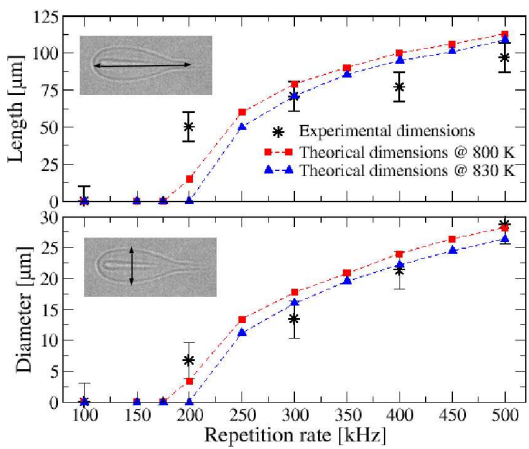}
 \caption{Quantitative comparison of experimental (stars) and modeling data (squares and triangles) versus RR. The dashed lines serve as guides to the eye.}
 \label{dim}
 \end{figure}

 For a more quantitative comparison, Fig.~\ref{dim} presents the measured and simulated dependence of the size (length and diameter) of the material modification versus the RR. For a threshold temperature near the strain point ($T_s=800$~K) or the annealing point ($T_a=830$~K), i.e.\ in the glass transition region $T_g$, we report good agreement between experiment and model. We did the same analysis for different numbers of pulses and pulse energies, and found the same agreement (not shown here). The fact that we find a threshold temperature around $T_g$ confirms that the observed modifications are mostly due to thermal accumulation, i.e. structural changes and related induced stress. For a RR lower than roughly 200 kHz, the predicted lattice temperature remains below $T_g$. The small modifications observed in the experiments at 100 and 200~kHz could thus be associated to non-thermal processes such as changes in the electronic structure. Actually, preliminary experimental results for the lowest RRs based on Difference Intensity Contrast microscopy exhibit slight modifications, corroborating non-thermal material modifications in this interaction regime.
 
In summary, a multi-scale model has been developed to describe permanent material modifications in dielectric materials irradiated by femtosecond pulse trains. Our model accounts for nonlinear laser pulse propagation as well as thermal diffusion. Simulations were performed for soda-lime glass, and directly confronted with experimental results. We report that depending of the RR, thermal accumulation in and around the focal volume leads to material modifications. It appears that the transition temperature of the glass can be considered as a good threshold for permanent material modifications. The size of the simulated modification volume is in good agreement with experimental observations in the thermal regime for RRs higher than 200~kHz. 
We expect that our model is applicable to a wide range of dielectrics. Thus, it will provide a new baseline for material structuring by fs lasers.

\begin{acknowledgments}
We acknowledge the European Commission, the French Ministry of Industry and the Aquitaine Regional Council for support and funding via the Femtoweld project. Numerical simulations were performed using computing
resources at M\'esocentre de Calcul Intensif Aquitain (MCIA) and Grand Equipement National pour le Calcul Intensif (GENCI, grant no.~2015-056129).
\end{acknowledgments}
\bibliography{article1}

\begin{thebibliography}{19}%
\makeatletter
\providecommand \@ifxundefined [1]{%
 \@ifx{#1\undefined}
}%
\providecommand \@ifnum [1]{%
 \ifnum #1\expandafter \@firstoftwo
 \else \expandafter \@secondoftwo
 \fi
}%
\providecommand \@ifx [1]{%
 \ifx #1\expandafter \@firstoftwo
 \else \expandafter \@secondoftwo
 \fi
}%
\providecommand \natexlab [1]{#1}%
\providecommand \enquote  [1]{``#1''}%
\providecommand \bibnamefont  [1]{#1}%
\providecommand \bibfnamefont [1]{#1}%
\providecommand \citenamefont [1]{#1}%
\providecommand \href@noop [0]{\@secondoftwo}%
\providecommand \href [0]{\begingroup \@sanitize@url \@href}%
\providecommand \@href[1]{\@@startlink{#1}\@@href}%
\providecommand \@@href[1]{\endgroup#1\@@endlink}%
\providecommand \@sanitize@url [0]{\catcode `\\12\catcode `\$12\catcode
  `\&12\catcode `\#12\catcode `\^12\catcode `\_12\catcode `\%12\relax}%
\providecommand \@@startlink[1]{}%
\providecommand \@@endlink[0]{}%
\providecommand \url  [0]{\begingroup\@sanitize@url \@url }%
\providecommand \@url [1]{\endgroup\@href {#1}{\urlprefix }}%
\providecommand \urlprefix  [0]{URL }%
\providecommand \Eprint [0]{\href }%
\providecommand \doibase [0]{http://dx.doi.org/}%
\providecommand \selectlanguage [0]{\@gobble}%
\providecommand \bibinfo  [0]{\@secondoftwo}%
\providecommand \bibfield  [0]{\@secondoftwo}%
\providecommand \translation [1]{[#1]}%
\providecommand \BibitemOpen [0]{}%
\providecommand \bibitemStop [0]{}%
\providecommand \bibitemNoStop [0]{.\EOS\space}%
\providecommand \EOS [0]{\spacefactor3000\relax}%
\providecommand \BibitemShut  [1]{\csname bibitem#1\endcsname}%
\let\auto@bib@innerbib\@empty
\bibitem [{\citenamefont {Miyamoto}\ \emph {et~al.}(2007)\citenamefont
  {Miyamoto}, \citenamefont {Horn}, \citenamefont {Gottmann}, \citenamefont
  {Wortmann},\ and\ \citenamefont {Yoshino}}]{miyamoto_fusion_2007}%
  \BibitemOpen
  \bibfield  {author} {\bibinfo {author} {\bibfnamefont {I.}~\bibnamefont
  {Miyamoto}}, \bibinfo {author} {\bibfnamefont {A.}~\bibnamefont {Horn}},
  \bibinfo {author} {\bibfnamefont {J.}~\bibnamefont {Gottmann}}, \bibinfo
  {author} {\bibfnamefont {D.}~\bibnamefont {Wortmann}}, \ and\ \bibinfo
  {author} {\bibfnamefont {F.}~\bibnamefont {Yoshino}},\ }\href {\doibase
  10.2961/jlmn.2007.01.0011} {\bibfield  {journal} {\bibinfo  {journal} {J.
  Laser Micro/Nanoeng.}\ }\textbf {\bibinfo {volume} {2}},\ \bibinfo {pages}
  {57} (\bibinfo {year} {2007})}\BibitemShut {NoStop}%
\bibitem [{\citenamefont {Peng}\ \emph {et~al.}(2013)\citenamefont {Peng},
  \citenamefont {Grojo}, \citenamefont {Rayner},\ and\ \citenamefont
  {Corkum}}]{peng_control_2013}%
  \BibitemOpen
  \bibfield  {author} {\bibinfo {author} {\bibfnamefont {J.}~\bibnamefont
  {Peng}}, \bibinfo {author} {\bibfnamefont {D.}~\bibnamefont {Grojo}},
  \bibinfo {author} {\bibfnamefont {D.~M.}\ \bibnamefont {Rayner}}, \ and\
  \bibinfo {author} {\bibfnamefont {P.~B.}\ \bibnamefont {Corkum}},\
  }\href@noop {} {\bibfield  {journal} {\bibinfo  {journal} {Appl. Phys.
  Lett.}\ }\textbf {\bibinfo {volume} {102}} (\bibinfo {year}
  {2013})}\BibitemShut {NoStop}%
\bibitem [{\citenamefont {Zhu}\ \emph {et~al.}(2005)\citenamefont {Zhu},
  \citenamefont {Van~Howe}, \citenamefont {Durst}, \citenamefont {Zipfel},\
  and\ \citenamefont {Xu}}]{Zhu20052153}%
  \BibitemOpen
  \bibfield  {author} {\bibinfo {author} {\bibfnamefont {G.}~\bibnamefont
  {Zhu}}, \bibinfo {author} {\bibfnamefont {J.}~\bibnamefont {Van~Howe}},
  \bibinfo {author} {\bibfnamefont {M.}~\bibnamefont {Durst}}, \bibinfo
  {author} {\bibfnamefont {W.}~\bibnamefont {Zipfel}}, \ and\ \bibinfo {author}
  {\bibfnamefont {C.}~\bibnamefont {Xu}},\ }\href {\doibase
  10.1364/OPEX.13.002153} {\bibfield  {journal} {\bibinfo  {journal} {Opt.
  Express}\ }\textbf {\bibinfo {volume} {13}},\ \bibinfo {pages} {2153}
  (\bibinfo {year} {2005})}\BibitemShut {NoStop}%
\bibitem [{\citenamefont {Cheng}\ \emph {et~al.}(2013)\citenamefont {Cheng},
  \citenamefont {Liu}, \citenamefont {Shang}, \citenamefont {Liu},
  \citenamefont {Perrie}, \citenamefont {Dearden},\ and\ \citenamefont
  {Watkins}}]{Cheng201388}%
  \BibitemOpen
  \bibfield  {author} {\bibinfo {author} {\bibfnamefont {J.}~\bibnamefont
  {Cheng}}, \bibinfo {author} {\bibfnamefont {C.-S.}\ \bibnamefont {Liu}},
  \bibinfo {author} {\bibfnamefont {S.}~\bibnamefont {Shang}}, \bibinfo
  {author} {\bibfnamefont {D.}~\bibnamefont {Liu}}, \bibinfo {author}
  {\bibfnamefont {W.}~\bibnamefont {Perrie}}, \bibinfo {author} {\bibfnamefont
  {G.}~\bibnamefont {Dearden}}, \ and\ \bibinfo {author} {\bibfnamefont
  {K.}~\bibnamefont {Watkins}},\ }\href {\doibase
  http://dx.doi.org/10.1016/j.optlastec.2012.06.037} {\bibfield  {journal}
  {\bibinfo  {journal} {Opt. Laser Technol.}\ }\textbf {\bibinfo {volume}
  {46}},\ \bibinfo {pages} {88 } (\bibinfo {year} {2013})}\BibitemShut
  {NoStop}%
\bibitem [{\citenamefont {Kymionis}\ \emph {et~al.}(2012)\citenamefont
  {Kymionis}, \citenamefont {Kankariya}, \citenamefont {Plaka},\ and\
  \citenamefont {Reinstein}}]{Kymionis}%
  \BibitemOpen
  \bibfield  {author} {\bibinfo {author} {\bibfnamefont {G.}~\bibnamefont
  {Kymionis}}, \bibinfo {author} {\bibfnamefont {V.}~\bibnamefont {Kankariya}},
  \bibinfo {author} {\bibfnamefont {A.}~\bibnamefont {Plaka}}, \ and\ \bibinfo
  {author} {\bibfnamefont {D.}~\bibnamefont {Reinstein}},\ }\href@noop {}
  {\bibfield  {journal} {\bibinfo  {journal} {J. Refract. Surg.}\ }\textbf
  {\bibinfo {volume} {28}},\ \bibinfo {pages} {912} (\bibinfo {year}
  {2012})}\BibitemShut {NoStop}%
\bibitem [{\citenamefont {Keldysh}(1965)}]{keldysh_ionization_1965}%
  \BibitemOpen
  \bibfield  {author} {\bibinfo {author} {\bibfnamefont {L.~V.}\ \bibnamefont
  {Keldysh}},\ }\href
  {http://www.slac.stanford.edu/grp/arb/tn/arbvol5/AARD451.pdf} {\bibfield
  {journal} {\bibinfo  {journal} {Soviet Physics {JETP}}\ }\textbf {\bibinfo
  {volume} {20}},\ \bibinfo {pages} {1307} (\bibinfo {year}
  {1965})}\BibitemShut {NoStop}%
\bibitem [{\citenamefont {Brouwer}\ and\ \citenamefont
  {Rethfeld}(2014)}]{brouwer_excitation_2014}%
  \BibitemOpen
  \bibfield  {author} {\bibinfo {author} {\bibfnamefont {N.}~\bibnamefont
  {Brouwer}}\ and\ \bibinfo {author} {\bibfnamefont {B.}~\bibnamefont
  {Rethfeld}},\ }\href@noop {} {\bibfield  {journal} {\bibinfo  {journal} {J.
  Opt. Soc. Am. B}\ }\textbf {\bibinfo {volume} {31}},\ \bibinfo {pages} {C28}
  (\bibinfo {year} {2014})}\BibitemShut {NoStop}%
\bibitem [{\citenamefont {Gamaly}\ and\ \citenamefont
  {Rode}(2014)}]{gamaly_transient_2014}%
  \BibitemOpen
  \bibfield  {author} {\bibinfo {author} {\bibfnamefont {E.~G.}\ \bibnamefont
  {Gamaly}}\ and\ \bibinfo {author} {\bibfnamefont {A.~V.}\ \bibnamefont
  {Rode}},\ }\href {\doibase 10.1364/JOSAB.31.000C36} {\bibfield  {journal}
  {\bibinfo  {journal} {J. Opt. Soc. Am. B}\ }\textbf {\bibinfo {volume}
  {31}},\ \bibinfo {pages} {C36} (\bibinfo {year} {2014})}\BibitemShut
  {NoStop}%
\bibitem [{\citenamefont {Burakov}\ \emph {et~al.}(2007)\citenamefont
  {Burakov}, \citenamefont {Bulgakova}, \citenamefont {Stoian}, \citenamefont
  {Mermillod-Blondin}, \citenamefont {Audouard}, \citenamefont {Rosenfeld},
  \citenamefont {Husakou},\ and\ \citenamefont
  {Hertel}}]{burakov_spatial_2007}%
  \BibitemOpen
  \bibfield  {author} {\bibinfo {author} {\bibfnamefont {I.~M.}\ \bibnamefont
  {Burakov}}, \bibinfo {author} {\bibfnamefont {N.~M.}\ \bibnamefont
  {Bulgakova}}, \bibinfo {author} {\bibfnamefont {R.}~\bibnamefont {Stoian}},
  \bibinfo {author} {\bibfnamefont {A.}~\bibnamefont {Mermillod-Blondin}},
  \bibinfo {author} {\bibfnamefont {E.}~\bibnamefont {Audouard}}, \bibinfo
  {author} {\bibfnamefont {A.}~\bibnamefont {Rosenfeld}}, \bibinfo {author}
  {\bibfnamefont {A.}~\bibnamefont {Husakou}}, \ and\ \bibinfo {author}
  {\bibfnamefont {I.~V.}\ \bibnamefont {Hertel}},\ }\href@noop {} {\bibfield
  {journal} {\bibinfo  {journal} {J. Appl. Phys.}\ }\textbf {\bibinfo {volume}
  {101}} (\bibinfo {year} {2007})}\BibitemShut {NoStop}%
\bibitem [{\citenamefont {Bulgakova}\ \emph {et~al.}(2014)\citenamefont
  {Bulgakova}, \citenamefont {Zhukov}, \citenamefont {Meshcheryakov},
  \citenamefont {Gemini}, \citenamefont {Brajer}, \citenamefont {Rostohar},\
  and\ \citenamefont {Mocek}}]{bulgakova_pulsed_2014}%
  \BibitemOpen
  \bibfield  {author} {\bibinfo {author} {\bibfnamefont {N.~M.}\ \bibnamefont
  {Bulgakova}}, \bibinfo {author} {\bibfnamefont {V.~P.}\ \bibnamefont
  {Zhukov}}, \bibinfo {author} {\bibfnamefont {Y.~P.}\ \bibnamefont
  {Meshcheryakov}}, \bibinfo {author} {\bibfnamefont {L.}~\bibnamefont
  {Gemini}}, \bibinfo {author} {\bibfnamefont {J.}~\bibnamefont {Brajer}},
  \bibinfo {author} {\bibfnamefont {D.}~\bibnamefont {Rostohar}}, \ and\
  \bibinfo {author} {\bibfnamefont {T.}~\bibnamefont {Mocek}},\ }\href@noop {}
  {\bibfield  {journal} {\bibinfo  {journal} {J. Opt. Soc. Am. B}\ }\textbf
  {\bibinfo {volume} {31}} (\bibinfo {year} {2014})}\BibitemShut {NoStop}%
\bibitem [{\citenamefont {Schaffer}, \citenamefont {Garcia},\ and\
  \citenamefont {Mazur}(2003)}]{Schaffer03}%
  \BibitemOpen
  \bibfield  {author} {\bibinfo {author} {\bibfnamefont {C.~B.}\ \bibnamefont
  {Schaffer}}, \bibinfo {author} {\bibfnamefont {J.~F.}\ \bibnamefont
  {Garcia}}, \ and\ \bibinfo {author} {\bibfnamefont {E.}~\bibnamefont
  {Mazur}},\ }\href@noop {} {\bibfield  {journal} {\bibinfo  {journal} {Appl.
  Phys. A}\ }\textbf {\bibinfo {volume} {76}},\ \bibinfo {pages} {351}
  (\bibinfo {year} {2003})}\BibitemShut {NoStop}%
\bibitem [{\citenamefont {Couairon}\ and\ \citenamefont
  {Mysyrowicz}(2007)}]{couairon_femtosecond_2007}%
  \BibitemOpen
  \bibfield  {author} {\bibinfo {author} {\bibfnamefont {A.}~\bibnamefont
  {Couairon}}\ and\ \bibinfo {author} {\bibfnamefont {A.}~\bibnamefont
  {Mysyrowicz}},\ }\href {\doibase 10.1016/j.physrep.2006.12.005} {\bibfield
  {journal} {\bibinfo  {journal} {Physics Reports}\ }\textbf {\bibinfo {volume}
  {441}},\ \bibinfo {pages} {47} (\bibinfo {year} {2007})}\BibitemShut
  {NoStop}%
\bibitem [{\citenamefont {Bulgakova}, \citenamefont {Stoian},\ and\
  \citenamefont {Rosenfeld}(2010)}]{bulgakova_laser-induced_2010}%
  \BibitemOpen
  \bibfield  {author} {\bibinfo {author} {\bibfnamefont {N.~M.}\ \bibnamefont
  {Bulgakova}}, \bibinfo {author} {\bibfnamefont {R.}~\bibnamefont {Stoian}}, \
  and\ \bibinfo {author} {\bibfnamefont {A.}~\bibnamefont {Rosenfeld}},\ }\href
  {http://iopscience.iop.org/1063-7818/40/11/A04} {\bibfield  {journal}
  {\bibinfo  {journal} {Quant. Electron.}\ }\textbf {\bibinfo {volume} {40}},\
  \bibinfo {pages} {966} (\bibinfo {year} {2010})}\BibitemShut {NoStop}%
\bibitem [{\citenamefont {Eaton}\ \emph {et~al.}(2005)\citenamefont {Eaton},
  \citenamefont {Zhang}, \citenamefont {Herman}, \citenamefont {Yoshino},
  \citenamefont {Shah}, \citenamefont {Bovatsek},\ and\ \citenamefont
  {Arai}}]{eaton_heat_2005}%
  \BibitemOpen
  \bibfield  {author} {\bibinfo {author} {\bibfnamefont {S.~M.}\ \bibnamefont
  {Eaton}}, \bibinfo {author} {\bibfnamefont {H.}~\bibnamefont {Zhang}},
  \bibinfo {author} {\bibfnamefont {P.~R.}\ \bibnamefont {Herman}}, \bibinfo
  {author} {\bibfnamefont {F.}~\bibnamefont {Yoshino}}, \bibinfo {author}
  {\bibfnamefont {L.}~\bibnamefont {Shah}}, \bibinfo {author} {\bibfnamefont
  {J.}~\bibnamefont {Bovatsek}}, \ and\ \bibinfo {author} {\bibfnamefont
  {A.~Y.}\ \bibnamefont {Arai}},\ }\href
  {http://www.opticsinfobase.org.ezproxy.univ-littoral.fr/DirectPDFAccess/F3EEED25-E0AA-6F25-1A5EE0AA48129BDF_84351/oe-13-12-4708.pdf?da=1&id=84351&seq=0&mobile=no}
  {\bibfield  {journal} {\bibinfo  {journal} {Opt. Express}\ }\textbf {\bibinfo
  {volume} {13}},\ \bibinfo {pages} {4708} (\bibinfo {year}
  {2005})}\BibitemShut {NoStop}%
\bibitem [{\citenamefont {Jamshidi-Ghaleh}, \citenamefont {Abdolahpour},\ and\
  \citenamefont {Mansour}(2006)}]{jamshidi-ghaleh_laser_2006}%
  \BibitemOpen
  \bibfield  {author} {\bibinfo {author} {\bibfnamefont {K.}~\bibnamefont
  {Jamshidi-Ghaleh}}, \bibinfo {author} {\bibfnamefont {D.}~\bibnamefont
  {Abdolahpour}}, \ and\ \bibinfo {author} {\bibfnamefont {N.}~\bibnamefont
  {Mansour}},\ }\href {\doibase 10.1002/lapl.200610056} {\bibfield  {journal}
  {\bibinfo  {journal} {Laser Phys. Lett.}\ }\textbf {\bibinfo {volume} {3}},\
  \bibinfo {pages} {573} (\bibinfo {year} {2006})}\BibitemShut {NoStop}%
\bibitem [{\citenamefont {Berg\'e}\ \emph {et~al.}(2007)\citenamefont
  {Berg\'e}, \citenamefont {Skupin}, \citenamefont {Nuter}, \citenamefont
  {Kasparian},\ and\ \citenamefont {Wolf}}]{berge_ultrashort_2007}%
  \BibitemOpen
  \bibfield  {author} {\bibinfo {author} {\bibfnamefont {L.}~\bibnamefont
  {Berg\'e}}, \bibinfo {author} {\bibfnamefont {S.}~\bibnamefont {Skupin}},
  \bibinfo {author} {\bibfnamefont {R.}~\bibnamefont {Nuter}}, \bibinfo
  {author} {\bibfnamefont {J.}~\bibnamefont {Kasparian}}, \ and\ \bibinfo
  {author} {\bibfnamefont {J.-P.}\ \bibnamefont {Wolf}},\ }\href {\doibase
  10.1088/0034-4885/70/10/R03} {\bibfield  {journal} {\bibinfo  {journal}
  {Reports on Progress in Physics}\ }\textbf {\bibinfo {volume} {70}},\
  \bibinfo {pages} {1633} (\bibinfo {year} {2007})}\BibitemShut {NoStop}%
\bibitem [{\citenamefont {Jewell}(1993)}]{jewell_thermooptic_1993}%
  \BibitemOpen
  \bibfield  {author} {\bibinfo {author} {\bibfnamefont {J.~M.}\ \bibnamefont
  {Jewell}},\ }\href {\doibase 10.1111/j.1151-2916.1993.tb06659.x} {\bibfield
  {journal} {\bibinfo  {journal} {J. Am. Ceram. Soc.}\ }\textbf {\bibinfo
  {volume} {76}},\ \bibinfo {pages} {1855} (\bibinfo {year}
  {1993})}\BibitemShut {NoStop}%
\bibitem [{\citenamefont {Sun}\ \emph {et~al.}(2006)\citenamefont {Sun},
  \citenamefont {Jiang}, \citenamefont {Liu}, \citenamefont {Wu}, \citenamefont
  {Yang},\ and\ \citenamefont {Gong}}]{sun_diagnose_2006}%
  \BibitemOpen
  \bibfield  {author} {\bibinfo {author} {\bibfnamefont {Q.}~\bibnamefont
  {Sun}}, \bibinfo {author} {\bibfnamefont {H.~B.}\ \bibnamefont {Jiang}},
  \bibinfo {author} {\bibfnamefont {Y.}~\bibnamefont {Liu}}, \bibinfo {author}
  {\bibfnamefont {Z.~X.}\ \bibnamefont {Wu}}, \bibinfo {author} {\bibfnamefont
  {H.}~\bibnamefont {Yang}}, \ and\ \bibinfo {author} {\bibfnamefont {Q.~H.}\
  \bibnamefont {Gong}},\ }\href
  {/home/caulier/Documents/Biblio/Organisée/Plasma/FrontPhysChina_1_Gong_2006_67.pdf}
  {\bibfield  {journal} {\bibinfo  {journal} {Frontiers of Physics in China}\
  }\textbf {\bibinfo {volume} {1}},\ \bibinfo {pages} {67} (\bibinfo {year}
  {2006})}\BibitemShut {NoStop}%
\bibitem [{\citenamefont {{Specialty Glass Products (SGP)}}()}]{sodalime}%
  \BibitemOpen
  \bibfield  {author} {\bibinfo {author} {\bibnamefont {{Specialty Glass
  Products (SGP)}}},\ }\href@noop {} {\enquote {\bibinfo {title} {{Soda Lime
  Product Sheet}},}\ }\bibinfo {howpublished}
  {\url{http://www.sgpinc.com/sodalime.htm}}\BibitemShut {NoStop}%
\end{thebibliography}%

\end{document}